# USING IMAGE PROCESSING TECHNIQUES TO INCREASE SAFETY IN SHOOTING RANGES

Ufuk Asıl and Efendi Nasibov

Department of Computer Science, Dokuz Eylul University, Izmir, Turkey

## ABSTRACT

Accidents are a leading cause of deaths in armed forces. The Aim of this paper is to minimize the accidents caused using weapons in the armed forces. Developing artificial intelligence technologies aim to increase efficiency more and more wherever people exist. Giving guns to inexperienced, untrained, or unpredictable mentally unhealthy people in shooting ranges used for gun training can be risky and fatal. With the use of image processing technologies in these shooting ranges, it is aimed to minimize the risk of life-threatening accidents that may be caused by this people. Artificial intelligence is trained for the targets to be used in shooting ranges. When the camera of weapon sees these targets, it switches from safe mode to firing mode. When a risky situation occurs in shooting range, the gun turns itself into safe mode with various additional security measures.

## 1. INTRODUCTION

A shooting range is a specialized facility designed for firearms qualifications, training or practicing as shown Figure 1. Some shooting ranges are operated by military or law enforcement agencies, though many ranges are privately owned and cater to recreational shooters. Each facility is typically overseen by one or more supervisory personnel, called a range master or "Range Safety Officer" (RSO) in the US, or a range conducting officer (RCO) in the UK. Supervisory personnel are responsible for ensuring that all weapon safety rules and relevant government regulations are followed at all times. [1]

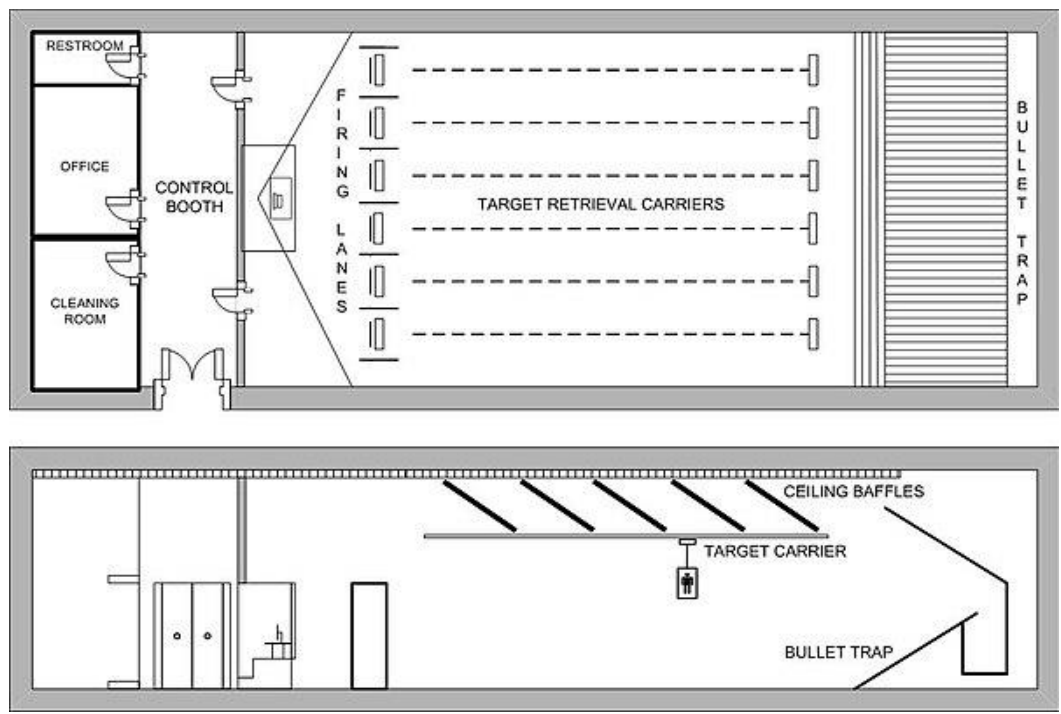

Figure 1.Floor and sectional diagrams of a typical indoor shooting range [2]

Shooting ranges can be outdoors as well as indoors. In the shooting ranges, usually there is one Security Chief, and when the security risk is high, a sub-security chief can be assigned to each shooter.

The main causes of human based accidents that may occur in the shooting ranges are as follows.

- There is a big difference between the trigger gap of an installed weapon and an uninstalled weapon. Many new shooters shoot involuntarily because they cannot understand this difference. Shooters receive theoretical training before shooting with guns, but the loud sound and gun recoil in the shooting ranges cause anxiety on shooters. Because of this anxiety, the shooter's muscles may show involuntary contractions. This may cause the shooter to press the trigger involuntarily and strike elsewhere outside the target.

- When the shooter does not understand the commands given by the shooting instructor, he tries to turn to make eye contact with the shooting instructor. Meanwhile, he can make involuntary deadly shots to his friends or shooting instructor by changing the direction from the target with the gun.

- In outdoor shooting ranges, other persons can enter between the target and the shooter unconsciously

- The hot bullet capsule that comes out of the gun may get into the trainee's clothes and cause mild burns. Burns causes some pain and pain is a good distracter; He will stop shooting and will try to remove the hot bullet shell that is his clothes with one hand, while having a gun ready to fire on the other; chances are he may involuntarily touch the trigger and shoot For this reason, the shooting of the rifle or gun outside the target involuntarily causes fatal accidents.

• Mental illnesses of individuals to be enlisted are difficult to diagnose in countries with compulsory military service. In addition, individuals who are exposed to a disciplined and isolated environment for the first time may have mental disorders later. These individuals can harm themselves in the training environment as well as harm the people around them [3].

Table 1. Mean – (min-max) and median of risk factors in Greek Army Forces [4]

| | min | max | Mean ±SD | Median |
|---|---|---|---|---|
| **\*Safety of the environment and equipment** | **18.0** | **40.0** | **25.4±3.7** | **\*24 (23 - 27)** |
| **\*Working environment** | **16.0** | **68.0** | **44.3±7.5** | **\*48 (37 - 50)** |
| Radiation | 0.0 | 8.0 | 5.7±2.4 | 6 (5 - 7) |
| Lighting | 0.0 | 20.0 | 12.4±2.6 | 13 (10 - 14) |
| Vibrations | 1.0 | 13.0 | 7.9±3.2 | 6 (5 - 12) |
| Electricity risks | 4.0 | 41.0 | 35.3±5,4 | 37 (33 - 39) |
| fire | 11.0 | 44.0 | 36±5.1 | 37 (35 - 39) |
| **\*Risk factors associated with the organization of work** | **3.0** | **12.0** | **7.3±1.8** | **\*7 (6 - 8)** |
| Hazards from equipment | 2.0 | 70.0 | 38.5±13.1 | 34 (29 - 51) |
| Risks from falling | 0,0 | 36.0 | 20.4±5.1 | 21 (19 - 21) |
| Flammable - chemicals | 4.0 | 62.0 | 43,1±12.6 | 42 (35 - 49) |
| Risks from manual handling | 0.0 | 8.0 | 5.2±0.9 | 5 (5 - 6) |
| Risks from the use of computers | 0.0 | 19.0 | 13±2.7 | 13 (12 - 14) |
| sounds | 2.0 | 12.0 | 7.1±1.3 | 7 (6 - 8) |
| Total Risk | 146.0 | 410.0 | 303±36.9 | 310 (289 - 325) |

As shown in Table 1 According to a research carried out in the Greek Army, the highest scores were seen in equipment and environmental safety issues. [4]

Many people may die or get injured because of unpredictable human-based errors in shooting ranges. Thanks to this technique we have developed, it is aimed to minimize human-based accidents in shooting areas where weapons modified with artificial intelligence technologies are used.

## 2. USED DEVICES AND FEATURES

### 2.1. Small AI Computer

Small computers have the necessary hardware for artificial intelligence applications, and they can run programming languages such as Python, C++. While these computers can be in general credit card size, some of them can be much smaller. The carrier board of small computer that allow interconnections other than the processor and graphics card can be reduced in size and redesigned with Printed circuit board (PCB) layout design. An example of this small computer is the Raspberry Pi[5], Jetson Nano [6], Jetson XavierNX [7], Intel Neural Compute Stick [8].

In this study, Jetson Nano small computer was used. Jetson Nano has a graphics processor. With help of this feature and parallel programing, it can process images very fast and high FPS values can be obtained. The FPS rates of various models given on Nvidia's official site are presented in Table2.There are just one or two small computers that faster than Jetson nano. These devices are used for professional purposes rather than education. Their cost is high. DNR (did not run) results given in the Table 2occurred frequently due to limited memory capacity, unsupported network layers, or hardware/software limitations.

Table 2. Deep Learning Inference Benchmarks [9]

| Model | Application | Framework | Jetson Nano | Raspberry Pi 3 | Intel NeuralComputeStick 2 |
|---|---|---|---|---|---|
| ResNet-50(224×224) | Classification | TensorFlow | 36 FPS | 1.4 FPS | 16 FPS |
| MobileNet-v2(300×300) | Classification | TensorFlow | 64 FPS | 2.5 FPS | 30 FPS |
| SSD ResNet-18(960×544) | Object Detection | TensorFlow | 5 FPS | DNR | DNR |
| SSD ResNet-18(480×272) | Object Detection | TensorFlow | 16 FPS | DNR | DNR |
| SSD ResNet-18(300×300) | Object Detection | TensorFlow | 18 FPS | DNR | DNR |
| SSD Mobilenet-V2(960×544) | Object Detection | TensorFlow | 8 FPS | DNR | 1.8 FPS |
| SSD Mobilenet-V2(480×272) | Object Detection | TensorFlow | 27 FPS | DNR | 7 FPS |
| SSD Mobilenet-V2(300×300) | Object Detection | TensorFlow | 39 FPS | 1 FPS | 11 FPS |
| Inception V4(299×299) | Classification | PyTorch | 11 FPS | DNR | DNR |
| Tiny YOLO V3(416×416) | Object Detection | Darknet | 25 FPS | 0.5 FPS | DNR |
| OpenPose(256×256) | PoseEstimation | Caffe | 14 FPS | DNR | 5 FPS |
| VGG-19(224×224) | Classification | MXNet | 10 FPS | 0.5 FPS | 5 FPS |
| SuperResolution(481×321) | Image Processing | PyTorch | 15 FPS | DNR | 0.6 FPS |
| Unet(1x512x512) | Segmentation | Caffe | 18 FPS | DNR | 5 FPS |

## 2.2. Servo Motor

A Servo motor is mounted on the safety pin of the rifle, that will be used to switch between modes like fire mode and safety mode. Servo motor is a linear actuator or rotary actuator that allows for precise control of linear or angular position, acceleration and velocity [10]. It consists of a motor coupled to a sensor for position feedback. It also requires a relatively sophisticated controller, often a dedicated module designed specifically for use with servomotors. Servo motors have three terminals. The two terminals are the electrical inputs required for the motor to operate. As shown in Figure 2, Other terminal controls the angular rotation required for switching.

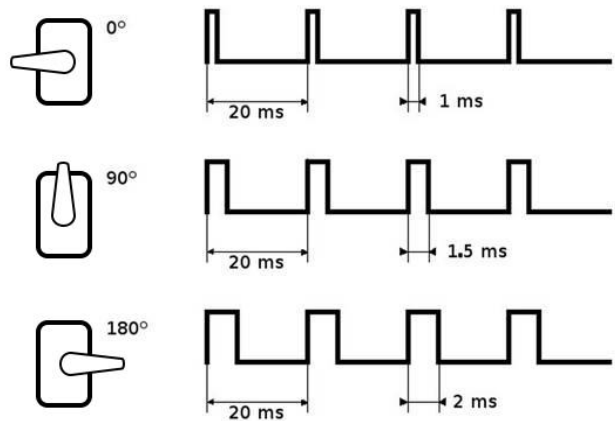

Figure 2. Variable pulse width control servo position

Communication of the desired position is performed by the transmission of a pulsed signal with a period of 20ms. The width of the pulse determines the position of the servo.

In general, on all models:

- A pulse between 500-1000 ms corresponds with 0º
- A pulse of 1500 ms corresponds with 90º (neutral point)
- A pulse between 2000-2500 ms corresponds to 180º

The servo used in this study has 4.8V/0.15sec / 60 deg Idle speed and its torque is 17.2kg.cm at 4.8V, but there are other faster and stronger professional servos on the market. As servo motor is an analog device, it does not work when directly attached to Jetson Nano. In order for the device to operate, there is a need for a PWM (Pulse Width Modulation) device that converts the data from analog todigital. AdafruidPWM (PulseWidth Modulation) was used in the study [11].

## 2.3. Camera

In this study, 12-megapixel High Quality Raspberry Pi Camera that was released in 2020 was used [12].All Raspberry Pi equipment compatible with jetson nano. There are many important reasons for using this camera module. Unlike the others, this camera has interchangeable lens feature. The interchangeable cs-mount lens structure of the camera module makes it possible to use the rifle at any distance and area. There are also companies that produce cameras for professional purposes. In order for the camera to be able to focus quickly at different distances, equipment such as a lidar, which can measure long distances and automatically change the focusing distance of the lens, can be employed [13].

## 3. DESIGN OF ARTIFICIAL INTELLIGENT RIFLE

In this study, MP5 rifle was used experimentally [14]. This rifle is more suitable than other weapons in terms of compatibility with servo motors. With MP5 rifles, the safety switch pin extends to the other side of the rifle. Therefore, the servo motor can be installed without disabling the safety pin. In this way, the user can disable the operating system at any time and switch to firing mode and continuous shooting mode. Also, the mp5 rifle is one of the most commonly used rifle in army forces in Turkey. The generalized blueprint of the MP5 rifle is shown in Figure 3. The usable empty parts of the rifle are written with an empty label. Figure 3 identifies the empty parts of the rifle and which components can be placed in these locations.

At first look, our rifle design is no different from a traditional weapon as appearance. it is a design that a person cannot predict that the weapon has an electronic structure. The empty areas of rifle are well utilized. This shows how useful and bright the future of our design is.

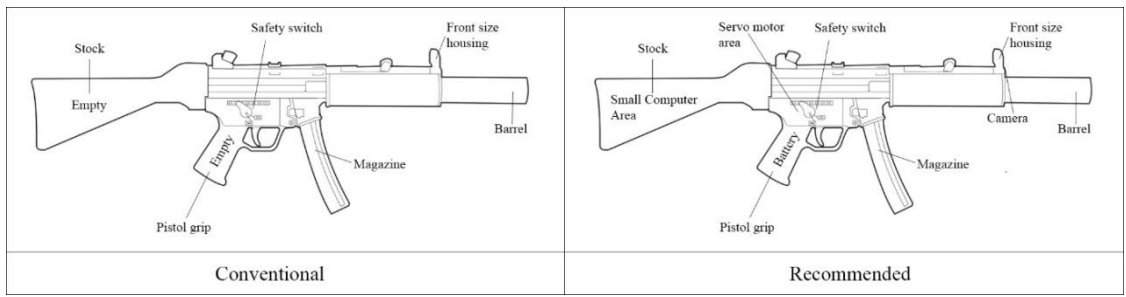

Figure 3. Recommended rifle illustration

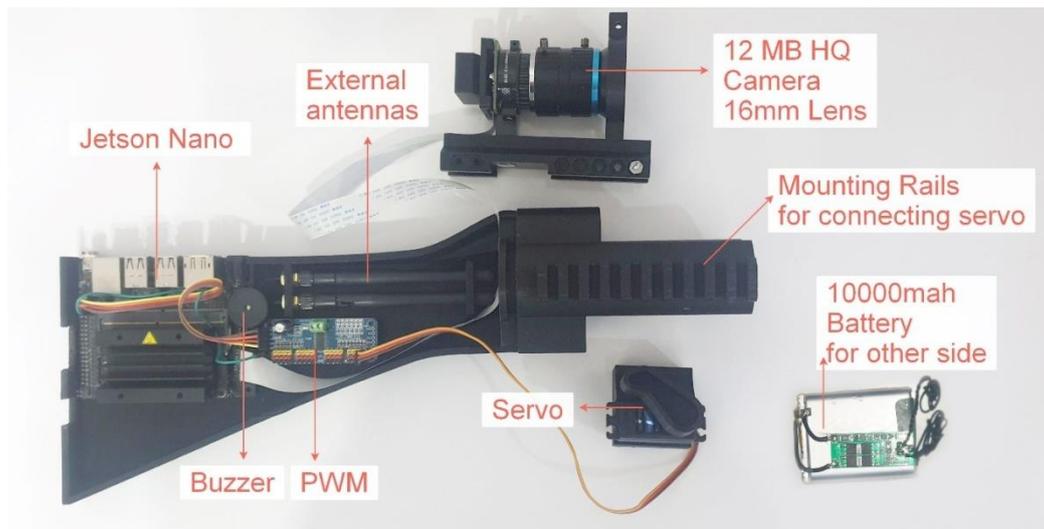

Figure 4. Real application of Ai mp5 rifle

The artificial intelligence conversion kit we have developed is shown in Figure 4. When the internal antennas are used, the area can be used more efficiently. The battery is placed on the other part of the stock. The change in weight of the rifle is almost negligible.

## 4. SOFTWARE ALGORITHMS AND PRINCIPLES

In this study, Python, Open CV (Open Source Computer Vision Library)[15]and SSD inception-V2[16]model was used for object detection. Open CV is a highly optimized open-source computer vision and machine learning library with focus on real-time applications.

SSD inception-V2 is a deep neural network used to detection objects in computer vision. It has two stages:

- It generates predictions about regions where an object can exist based on the input image.
- It predicts the class of the object, refines the bounding box and it does not create a pixel-level mask of the object like Mask R-CNN [17] algorithm

Transfer learning is a machine learning method where a model developed for a task is reused as the starting point for a model on a second task [18].In this way, efficient neural networks can be built even with 400-500 photos. We created two classes named "person" and "target" by retraining our model using transfer learning.

The main stages of the algorithm are as follows as shown in Figure 5

1. As the initial state Security pin is secured on Safety mode
2. If the camera sees a target, the rifle will be ready to fire with the security pin by coming to the shooting mode as in Figure 6.
3. Even if camera sees a target, there is a person on this frame, the rifle will be in off-shot mode with security pin by coming to the Safety mode as in Figure 7.

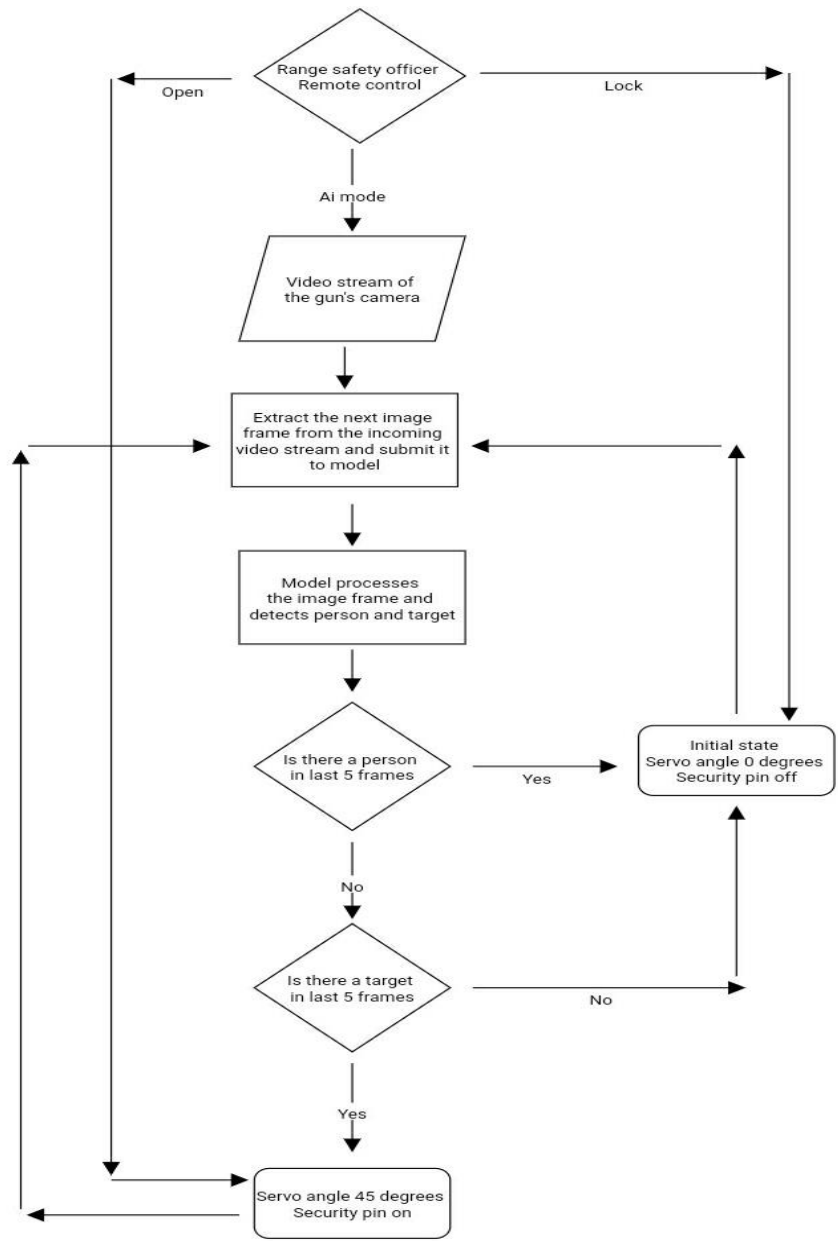

Figure 5. Flowchart of algorithm

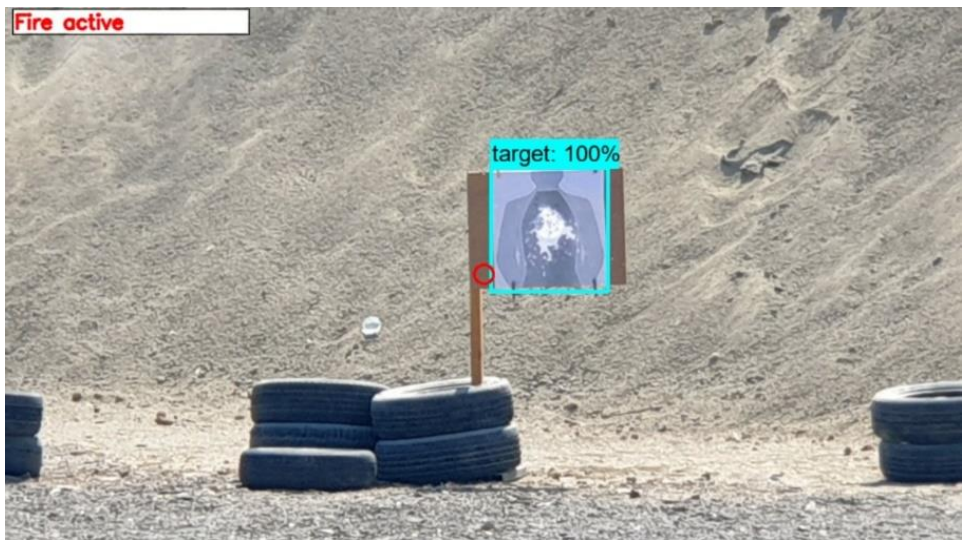

Figure 6. Security pin on fire mode

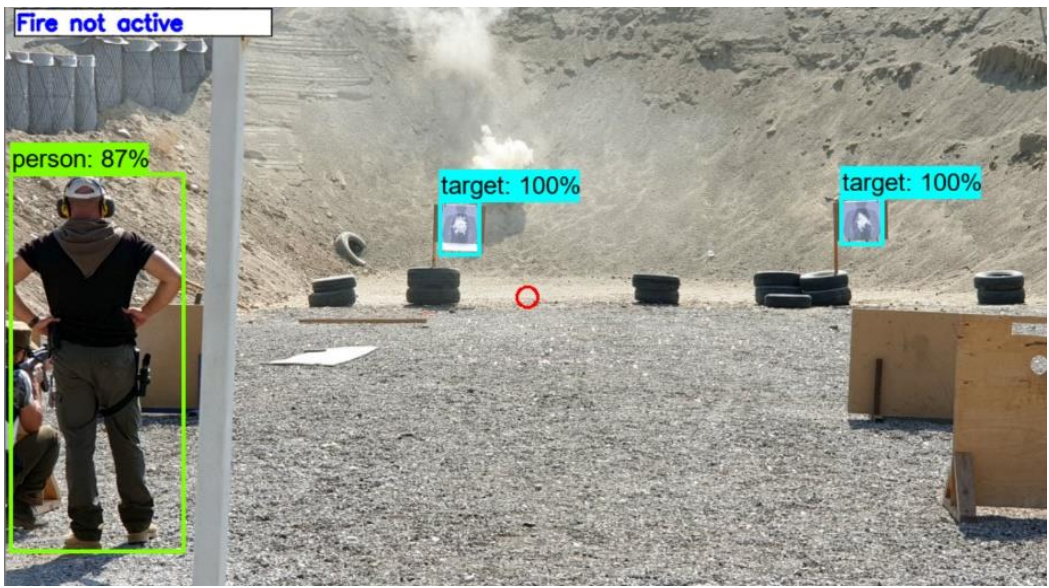

Figure 7. Security pin on safety mode

When the rifle sees the target, it commands the servo motor to switch shooting mode and tells the shooter that he is ready to fire with continuous sound from the Buzzer Speaker. The shooter can understand that the gun is ready to fire with a very low-cost equipment without the need for visual warning, and when it can be used without checking the security pin.

In addition to this technique,

- All weapons can be deactivated, activated by remote control.
- All weapons can be connected to a single center from where statistics and information can be kept.
- The success of the shooter can be evaluated in training.

# 5. CONCLUSION

In this study, it has been investigated whether an Artificial Intelligence image processing technology can reduce the security risk in shooting ranges. In all cases where our algorithm does not see the target keeping the firearm in safe mode reduces the risk of lack of vision due to the focus and movement of the camera to zero when switching from safety mode to fire mode. Risk is always present as it takes some time to switch to safe mode while the gun is in firing mode. Therefore, using devices such as Xavier NX which can be used in professional areas with high image processing features, will reduce the security risk to a minimum. These devices are not 100% safe and should only be used to increase security. With the SSD inception-V2 algorithm with approximately 20 FPS might have 0,05sec delay, servo motor with 0,10 sec delay and camera with 0.12 sec delay, and at least 0.27 seconds of total system delay was achieved. Thanks to devices such as Xavier NX and specially produced fast servo motors and no delay cameras, it will be able to react faster than a human being with large accuracy in results. A weapon programmed with this technique can be programmed with many conditions such as preventing a shot on a police officer or, a shot on a child or a shot on head, or a shot on someone without a gun.

## AUTHORS

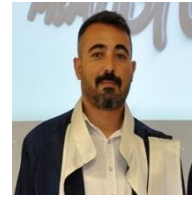

**Ufuk Asıl** is currently Ph.D. student of Department of Computer Science, Dokuz Eylul University, Izmir, Turkey. He is a Police officer since 2010. Hereceived his master's degree in Nanoscience and Nanoengineering from Dokuz Eylul University. His research interests Image Processing and Neuromorphic chip design.

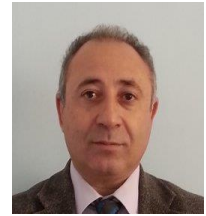

**Efendi Nasibov** (Nasiboglu) received the B.Sc. and M.Sc. Degrees in Applied Mathematics Department from Baku State University, Azerbaijan, and the Ph.D. degree in Mathematical Cybernetics (Moscow) and Dr.Sc. in Computer Science degree from Institute of Cybernetics of Academy of Science of Azerbaijan. He is currently a full Professor of Department of Computer Science, Dokuz Eylul University, Izmir, Turkey. His research interests are in the application of Fuzzy Modeling, Data Mining and AI techniques in Decision Making problems.